\documentclass{amsart}

\usepackage{url}
\usepackage{amsmath,amssymb,amsfonts}
\usepackage{graphicx,epsfig,epsf}
\usepackage{cite}
\usepackage{listings}
\usepackage{subfig}

\theoremstyle{definition}

\newtheorem{theorem}{Theorem}[section]

\newtheorem{corollary}{Corollary}[section]
\newtheorem{definition}{Definition}[section]
\newtheorem{example}{Example}[section]

\title{Description of the Chord Protocol using ASMs Formalism}

\author{Bojan Marinkovi\'{c}}
\address{Mathematical Institute of the Serbian Academy of Sciences and Arts, Belgrade, Serbia}
\email{\url{bojanm@mi.sanu.ac.rs}}

\author{Paola Glavan}
\address{Department of Mathematics and Descriptive Geometry,Faculty of Mechanical Engineering and Naval Architecture, Zagreb, Croatia}

\author{Zoran Ognjanovi\'{c}} 
\address{Mathematical Institute of the Serbian Academy of Sciences and Arts, Belgrade, Serbia}

\begin{document}
\maketitle

\begin{abstract}
This paper describes the overlay protocol Chord using the
formalism of Abstract State Machines. The formalization concerns
Chord actions that maintain ring topology and manipulate
distributed keys. We define a class of runs and prove the
correctness of our formalization with respect to it.

\end{abstract}

\section{Introduction}

A decentralized Peer-to-Peer system (P2P) \cite{P2P2010} involves many peers (nodes) which execute the same software, participate in the system
having equal rights and might join or leave the system continuously. In such a framework processes are dynamically distributed to
peers, with no centralized control. P2P systems have no inherent bottlenecks and can potentially scale very well. Moreover, since
there are no dedicated nodes critical for systems' functioning, those systems are resilient to failures, attacks, etc. The main
applications of P2P-systems involve: file sharing, redundant storage, real-time media streaming, etc.

P2P systems are frequently implemented in a form of overlay networks \cite{p2p-book}, a structure that is totally independent of
the underlying network that is actually connecting devices. Overlay network represents a logical look on organization of the
resources.
Some of the overlay networks are realized in the form of Distributed Hash Tables (DHT) that provide a lookup service similar to a hash
table; (key, value) pairs are stored in a DHT, and any participating peer can efficiently retrieve the value associated with a given
key. Responsibility for maintaining the mapping from keys to values is distributed among the peers, in such a way that any change in
the set of participants causes a minimal amount of disruption. It allows a DHT to scale to extremely large number of peers and to
handle continual node arrivals, departures, and failures.
The Chord protocol \cite{Chord,Chord-TR,Chord-IEEE} is one of the first, simplest and most popular DHTs.
The paper \cite{Chord} which introduces Chord has been recently awarded the SIGCOMM 2011 Test-of-Time Award.

Our aim is to describe Chord using Abstract State Machine (ASM)
\cite{GUR95} and to prove the correctness of the formalization,
which was motivated by the obvious fact that errors in concurrent
systems are difficult to reproduce and find merely by program
testing. There are at least two reasons for using ASMs. 
The ASM-code for Chord presented in this paper has been written
following one of the best implementations \cite{Bamberg} of the
high level C$++$-like pseudo code from \cite{Chord-IEEE}.

Recently, several non-relational database systems (NRDBMS) have
been developed \cite{LaM10, bigtable} that are usually based on
the Chord like technology. To analyze their behavior, it might be
useful to characterize situations when correctness of the
underlying protocol holds. Following that idea, we have formulated
several deterministic conditions that guarantee correctness of
Chord, and proved the corresponding statements. This is in
contrast to the approach from
\cite{Chord,Chord-TR,Chord-IEEE,liben02} where a probabilistic
analysis is proposed, and correctness holds with "high
probability".

 Most of them are based on the Chord like technology. NRDBMSs are
 used when a large amount of data exist and do not need frequent
 update. Usually, an NRDBMS does not guarantee correctness.
 
The main objectives of Chord are maintaining the ring topology as
nodes concurrently join and leave a network, mapping keys onto
nodes and distributed data handling.
The formalism of ASM enables us to precisely describe a class of
possible runs - so called regular runs - of the protocol, and to
prove correctness of the main operations with respect to it.
Moreover, several examples of runs, given in Example
\ref{primer_rr}, that violate the constraints for the regular runs
illustrate how correctness can be broken in those cases.

\section{ASM Formalization of Chord}
\label{formal_chord}

\subsection{Basic Notions}

Let $L$, $M$ and $K$ be three fixed positive integers, and $N = 2^M$. We will consider the following disjoint universes:
\begin{itemize}
    \item the set $Peer = \{p_1, \ldots, p_L \}$ of all peers that might participate in the considered Chord network,

 %   \item the set $IPs$ containing $L$ distinct IP-addresses of the elements of the set $Peers$,

    \item the set $Key = \{ k_1, \ldots, k_K\}$ of identifiers of objects that might be stored in the considered Chord network, and
    the set $Value = \{ v_1, \ldots, v_K\}$ of the values of those $K$ objects,

    \item the set $Chord = \{0, 1, \ldots , N-1\}$ denoting at most $N$ peers that are involved in the network in a particular
    moment,

    \item the sets $Join = \{join, skip\}$ and $Action=\{put, get, fair\_leave, unfair\_leave, skip\}$  which represent the actions of the
    peers.
\end{itemize}
Note that:
\begin{itemize}
    \item it might be that $L > N$ ($K > N$), i.e., that there are more peers (objects to be stored in the network) than nodes,
    but it can never be $N > L$, and

    \item without any loss of generality we assume that the numbers of keys and values are the same; if there are more values than
    keys, all values mapped to the same key might be organized in a list.
\end{itemize}

\begin{figure}
\centering
\includegraphics[width=0.33\textwidth]{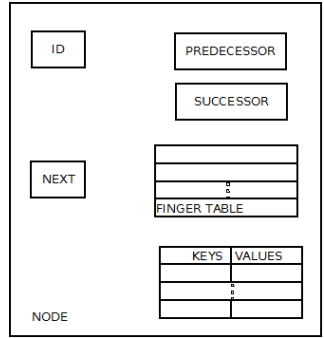}

\caption{Structure of Chord node} \label{node}
%\vspace{-0.5cm}
\end{figure}

Any peer, active in the network
%(i.e., associated with some $x \in Chord$, such that $Chord_f(x) = true$)
will be called a {\em node}.
We assume that a $node$ (Fig. \ref{node}) is represented by its identifier $id$ in the network, information on its $predecessor$
 and $successor$, a $finger$  {\em table}, a pointer ($next$) to an element in the finger table which will be updated
in the current stabilization cycle, and a {\em list of $\langle key, value \rangle$ pairs} of the records for which the node is responsible for.
% The sets of all keys, the corresponding
% values, and IP addresses of nodes are denoted by $Keys$, $Values$, and $IPs$, respectively.

More formally, we introduce the following functions:
\begin{itemize}
    \item $id: Peer \rightarrow Chord \cup \{ undef \}$
    \item $successor : Chord \rightarrow Chord$,
    \item $predecessor: Chord \rightarrow Chord$,
    \item $finger: Chord \rightarrow Chord^*$,
    \item $next:  Chord \rightarrow \{1, \ldots, M\}$, and
    \item $keyvalue: Chord \rightarrow (Chord \times Value)^*$,
\end{itemize}
where $Chord^*$ is the set that contains lists of nodes' identifiers, and $(Chord \times Value)^*$ is the set of lists
containing pairs $\langle hash(key), value \rangle$. Each $finger(x)$ has $M$ entries ordered respect to the ring ordering.

In other words, a peer $p$, which is a node, is represented by the tuple:
\begin{itemize}
    \item $\langle id(p), successor(id(p)), predecessor(id(p)), finger(id(p)), next(id(p)), keyvalue(id(p)) \rangle$.
\end{itemize}

Table \ref{chord_functions} shows all the other functions that
will be used in the formal description of the protocol, but that
do not directly change the representations of nodes. We assume
that the five functions in Table \ref{chord_functions} ($hash$,
$ping$,
%\\$nonderministic\_choice$,$communicaton$,
$known\_nodes$, $key\_value$ and $keys$) are external.

\begin{table}
\centering
\begin{tabular}[]{|l|l|}
\hline \textbf{Function} & \textbf{Description}  \\
\hline
$hash$ & Maps the sets of peers and keys to $Chord \cup \{ undef \}$\\
\hline
%
% $id$ & Maps the set of peers to $Chord \cup \{ undef \}$\\
%
% \hline
%
$ping$ & Tests whether a node is reachable \\
\hline
%
%$nonderministic\_choice$ & Simulates peers' nondeterministic choice of actions\\
%
%\hline
%
$member\_of$ & Checks whether a node is between two nodes in $Chord$  \\
\hline
$communication$ & Realizes communication requests \\
\hline
%
%$find\_successor$ &  Returns successor of a node or a responsible node for a given key\\
%
%\hline
%
%$get$ & Returns the value for the given key if it is stored in the network \\
%
%\hline
%
$mode$ & Determines $Peer\_agent$ state \\
\hline
$known\_nodes$ & Simulates external knowledge about existing nodes \\
\hline
$key\_value$ & Select a $\langle key, value\rangle$ for storing in the network\\
\hline
$keys$ & Select to look for a $value$ with particular $key$\\
\hline
\end{tabular}

\caption{Chord functions} \label{chord_functions}
% \vspace{-0.5cm}
\end{table}

The {\em hash} function assigns identifiers of nodes to peers and keys:
\begin{itemize}
    \item $hash: Peer \cup Key  \rightarrow Chord \cup \{undef\}$,
\end{itemize}
where {\em undef} is a special value which indicates that:
\begin{itemize}
    \item there are $N$ nodes in the network, and an identifier is requested for a new node, or
    \item there are $N$ keys in the network, but we try to add a new key.
\end{itemize}
The function must also guarantee that in each moment two different active peers (keys) have different hash values. However, note that
it is possible that in different moments different peers have the same identifier. Also, it may happen that a peer can have different
identifiers (obtained by different calls of the $hash$ function before and after a period in which the peer is not present in the
network).
The above mentioned {\em id} function can be explained as a "local" counterpart of {\em hash}. Namely, we can assume that the values
produced by {\em hash} are stored in the local memory and read and published by {\em id} to reduce the number of expensive calls of
{\em hash}.
In the program given below, $id$ will be invoked with the argument $Me$ to allow a node to identify itself in the network.

The external function {\em ping}, defined as:
\begin{itemize}
    \item $ping: Chord \rightarrow \{true, false\}$
\end{itemize}
returns $true$ or $false$, depending on whether the argument is reachable in the network.
% is no communication problem with $Node_{id}$ for the given $id$.

% In the explanation of the next two functions we will use the relations $\leqslant$ and $<$ which respect the ring ordering, i.e., that
% $0$ is the first successor of $N-1$.

%The external function {\em nondeterministic\_choice}:
%\begin{itemize}
%    \item $nondeterministic\_choice: Peers \rightarrow Actions$
%\end{itemize}
%simulates peers' nondeterministic choice of actions.

The function {\em member$\_$of}:
\begin{itemize}
    \item $member\_of: Chord \times Chord \times Chord \rightarrow \{true, false\}$,
\end{itemize}
determines whether the first argument is between two next two arguments with respect to the ring ordering,
more formally:
\begin{itemize}
  \item if $arg_2 = arg_3$ always returns $true$,
  \item if $arg_2 < arg_3$ returns $true$ if $arg_2 < arg_1 \leqslant arg_3$ holds,
  \item if $arg_2 > arg_3$ returns $true$ if $\neg (arg_3 \leqslant arg_1 < arg_2)$ holds,
  \item otherwise returns $false$.
\end{itemize}

The function $mode$:
\begin{itemize}
\item $mode: Peer \rightarrow Mode$
\end{itemize}
determines $Peer\_agent$ state. Initially, for all $p \in Peer$ value of $mode(p)$ is set to $not\_connected$.

% The function {\em loop\_type}:
% \begin{itemize}
% \item $loop\_type: Peer \rightarrow Action$
% \end{itemize}
% determines what action the $Peer\_agent$ will perform. If the $loop\_type(p) \notin \{ put, get, fair\_leave, unfair\_leave \}$ for some
% $p \in Peer$, peer $p$ will not do anything.

The external function $known\_nodes$:
\begin{itemize}
  \item $known\_nodes: Peer \rightarrow Chord$
\end{itemize}
simulates external knowledge about the nodes in the particular Chord network.

The external function $key\_value$:
\begin{itemize}
  \item $key\_value: Peer \rightarrow Key \times Value$
\end{itemize}
simulates the choice of a node to store a $\langle key,
value\rangle$ pair in the Chord network.

The external function $keys$:
\begin{itemize}
  \item $keys: Peer \rightarrow Key$
\end{itemize}
simulates the choice of a node to look if some $value$ with particular $key$ is stored in the Chord network.

\subsection{Chord Rules}
\label{peer_modul}

The rest of this section contains our ASM-formalization of the
Chord protocol. We present the general program executed by every
peer, and a high level description of the rules performed in a
Chord network which corresponds to the pseudo code given in
\cite{Chord-IEEE} (note that the rules \textsc{FairLeave},
\textsc{UnfairLeave}, \textsc{Put} and \textsc{Get} are not given
there). A detailed specification of these rules is provided in
Appendix \ref{chord_rules_details}.

\subsubsection{Peer\_agent Module}

The following main module contains actions that are executed by
every peer. The mode of all peers is initially {\em
not\_connected}. After a node joins a network successfully, its
mode is changed to {\em connected}. In each execution of a loop, a
node concurrently calls the rules responsible for the ring
topology maintenance (\textsc{Stabilize},
\textsc{UpdatePredecessor}, \textsc{UpdateFingers}) and
communication (\textsc{ReadMessages}) and, according to a
non-deterministic choice, it might also invoke one of the
\textsc{FairLeave}, \textsc{UnfairLeave}, \textsc{Put} and
\textsc{Get} rules.

\begin{lstlisting}[escapechar=\%]
if %$mode(Me) = not\_connected$% then
  if Choosed Action Is Join
    seq
      if There Are No Known Nodes then
        %\textsc{Start}%
      else
        %\textsc{Join}%
      endif
      if Connection Successful then
        %$mode(Me) := connected$%
      else
        %$mode(Me) := not\_connected$%
      endif
    endseq
  endif
else
  if %$mode(Me) = connected$% then
    if %$id(Me)$% Does Not Have Communication Problems then
      par
        %\textsc{Stabilize}%
        %\textsc{UpdatePredecessor}%
        %\textsc{UpdateFingers}%
          seq
            choose action in Action
            par
              %$LeavingActions =$%
                %\textsc{FairLeave}% Or %\textsc{UnfairLeave}%
              %$KeyValueHandling =$%
                %\textsc{Put}% Or %\textsc{Get}%
            endpar
          endseq
      endpar
    else
      %$mode(Me) := not\_connected$%
    endif
  endif
endif
\end{lstlisting}

\subsubsection{Chord Rules - High Level Description}

\begin{table}
\centering
 \begin{tabular}[]{|l|l|l|}
  \hline
  \textbf{Rule} & \textbf{Description} & \textbf{Resulting State} \\
  \hline
  \textsc{Start} & The first node starts the network & A state with one node \\
  %
  %\hline
  %$End$ & The last node leaves the network & "Empty" state $S_{\varepsilon}$ \\
  %
  \hline
  \textsc{Join} & A new node joins the network & A state with an additional \\
   & & node \\
  \hline
  \textsc{FairLeave} & A node leaves fairly the network & A state without one node \\
  \hline
  \textsc{UnfairLeave} & A node leaves/crashes & A state without one node \\
  \hline
  \textsc{Stabilize} & A node updates its successor & Successor and predecessor\\
   & and predecessor & update\\
  \hline
  \textsc{UpdatePredecessor} & Periodic check of the predecessor & Predecessor update \\
  \hline
  \textsc{UpdateFingers} & A node runs update on its  & Updating finger table entries  \\
  & finger table& \\
  \hline
  \textsc{Put} & A new (key, value) pair is stored & Updating (key,value) table \\
  \hline
  \textsc{Get} & Finding a value for a given key & Unchanged state \\ \hline
  \textsc{FindSuccessor} & Finding a responsible node for & Unchanged state \\
  & given key or successor of a node& \\
  \hline
  \textsc{ReadMessage} & Read messages dedicated to a node & Changing some local  \\
  & & variables if it is requested \\ \hline
  %%
  %\textsc{WaitForResponse} & Wait for the answer from the  & Unchanged state \\
  %& other node& \\ \hline
\end{tabular}

\caption{Chord rules} \label{chord_actions}
\vspace{-0.5cm}
\end{table}

Any node present in a Chord network can execute \textsc{Get} rule (ask for the value of a key). That rule does not
change the actual state of the network, but we define it as:
\begin{lstlisting}[escapechar=\%]
%\textsc{Get}$=$%
Invoke %\textsc{FindSuccessor}% For Given %$key$%
And Check Corresponding %$value$%
\end{lstlisting}

During the each execution of a {\em Peer\_agent Module} all the messages send to a node are processed:
\begin{lstlisting}[escapechar=\%]
%\textsc{ReadMessages}$=$%
Read Messages Dedicated To %$Me$%,
Change Local Variables If It Is Requested And
Clear Processed Messages
\end{lstlisting}

\section{Correctness of the Formalization}
\label{dokazi}

In this section we present the correctness of our formalization with respect to the so-called regular runs.

\begin{definition}
Let $x_1, x_2 \in Node$ and $y_0, \ldots y_r \in Node$ be all the
nodes from a Chord network such that $x_1 = y_0 < \ldots < y_r =
x_2$. The pair $\langle x_1, x_2 \rangle$ forms a {\em stable
pair} in a state if the following holds:
\begin{itemize}
    %\item $x_1=predecessor(y_0)$, $x_2=successor(y_r)$, and
    \item $y_{i+1} = successor (y_i)$, $y_{i} = predecessor (y_{i+1})$, for all $i \in \{ 0,\ldots ,r-1 \}$.
    %\item there is no $x_3 \in Node$, such that $member\_of(x_3,x_1,x_2) = true$ and $x_2=successor(x_3)$.
\end{itemize}
A Chord network $\{x_0, \ldots , x_{k-1} \}$, $k \geqslant 1,$ is
{\em stable} in a state if the pair $\langle x_0, x_0 \rangle$ is
stable. \hfill $\square$
\end{definition}
Intuitively, a pair $\langle x_1, x_2 \rangle$ is stable in a
state if there is no node trying to join the network through the
node on the ring-interval $(x_1,x_2)$ in that state.

\begin{definition}
{\em Regular runs} are all runs of a distributive algebra
$\mathcal{A}$
% . In the sequel we will restrict our attention to a subclass of regular runs, called {\em regular runs} which additionally
which satisfy that:
\begin{itemize}
%    \item all actions (i.e., the rules defined in Section \ref{chord_rules_details}) are atomic, \label{ogranicenje1}

    \item any execution of \textsc{FairLeave}, \textsc{UnfairLeave} and \textsc{Put} might happen only between a stable pair of nodes.
    \label{ogranicenje1}\hfill $\square$

%    \item a node cannot try to join a network via another node which leaves the network. \label{ogranicenje2}
%   nepotrebno zbog atomicnosti modula
%
%    \item the fairness condition, in the sense that every node will eventually execute its next move. \label{ogranicenje3}
\end{itemize}
\end{definition}

The following example illustrates the need for the above constraint. In the example and in the rest of the paper we will graphically
illustrate sequences of moves, so that $S_i$ denotes a state, the updated values are in bold, and $\diamondsuit$ means that the rest
of a network is not affected by a move.

\begin{example} \label{primer_rr}
Let $S_0$ be the initial state in which the nodes $N_1$ and $N_3$
are members of a network, and
the node $N_2$ wants to join. Suppose that before the pair
$\langle N_1, N_3 \rangle$ becomes stable, $N_1$ executes the put
rule with the hash $2$ of a key. Since $N_1$ is not aware of
$N_2$, the corresponding key will be stored in $N_3$, and not in
$N_2$.
\begin{center}
\begin{tabular}{|c|c|c|}
\hline
\multicolumn{3}{|c|}{\boldmath{$S_0$}} \\
\hline
$id$ & $1$ & $3$\\
\hline
$predecessor$ & $\diamondsuit$ & $1$ \\
\hline
$successor$ & $3$ & $\diamondsuit$ \\
\hline
$hash(key)$ & $empty$ & $empty$ \\
\hline
\end{tabular}
$\xrightarrow{N_{2} \textsc{Join}}$
\begin{tabular}{|c|c|c|c|}
\hline
\multicolumn{4}{|c|}{\boldmath{$S_1$}} \\
\hline
$id$ & $1$ & \boldmath{$2$} & $3$\\
\hline
$predecessor$ & $\diamondsuit$ & \boldmath{$undef$} & $1$ \\
\hline
$successor$ & $3$ & \boldmath{$3$} & $\diamondsuit$ \\
\hline
$hash(key)$ & $empty$ & \boldmath{$empty$} & $empty$ \\
\hline
\end{tabular}
\vspace{1mm}
$\xrightarrow{N_{2} \textsc{Stabilize}}$
\begin{tabular}{|c|c|c|c|}
\hline
\multicolumn{4}{|c|}{\boldmath{$S_2$}} \\
\hline
$id$ & $1$ & $2$ & $3$\\
\hline
$predecessor$ & $\diamondsuit$ & $undef$ & \boldmath{$2$} \\
\hline
$successor$ & $3$ & $3$ & $\diamondsuit$ \\
\hline
$hash(key)$ & $empty$ & $empty$ & $empty$ \\
\hline
\end{tabular}
$\xrightarrow{N_{1} \textsc{Put}(Key 2)}$
\begin{tabular}{|c|c|c|c|}
\hline
\multicolumn{4}{|c|}{\boldmath{$S_3$}} \\
\hline
$id$ & $1$ & $2$ & $3$\\
\hline
$predecessor$ & $\diamondsuit$ & $undef$ & $2$ \\
\hline
$successor$ & $3$ & $3$ & $\diamondsuit$ \\
\hline
$hash(key)$ & $empty$ & $empty$ & \boldmath{$2$} \\
\hline
\end{tabular}
\vspace{1mm}
$\xrightarrow{N_{1} \textsc{Stabilize}}$
\begin{tabular}{|c|c|c|c|}
\hline
\multicolumn{4}{|c|}{\boldmath{$S_4$}} \\
\hline
$id$ & $1$ & $2$ & $3$\\
\hline
$predecessor$ & $\diamondsuit$ & $undef$ & $2$ \\
\hline
$successor$ & \boldmath{$2$} & $3$ & $\diamondsuit$ \\
\hline
$hash(key)$ & $empty$ & $empty$ & $2$ \\
\hline
\end{tabular}
$\xrightarrow{N_{1} \textsc{Stabilize}}$
\begin{tabular}{|c|c|c|c|}
\hline
\multicolumn{4}{|c|}{\boldmath{$S_5$}} \\
\hline
$id$ & $1$ & $2$ & $3$\\
\hline
$predecessor$ & $\diamondsuit$ & \boldmath{$1$} & $2$ \\
\hline
$successor$ & $2$ & $3$ & $\diamondsuit$ \\
\hline
$hash(key)$ & $empty$ & $empty$ & $2$ \\
\hline
\end{tabular}
\end{center}

Again, assume that $S_0$ is the initial state and the network contains the nodes $N_1$, $N_3$ and $N_4$.
If the node $N_2$ executes the join rule, and before the pair
$\langle N_1, N_4 \rangle$ becomes stable, $N_3$ wants to leave,
$N_2$ will be isolated from the rest of the network, and the other
nodes will  never be aware of it.

\begin{center}
\begin{tabular}{|c|c|c|c|}
\hline
\multicolumn{4}{|c|}{\boldmath{$S_0$}} \\
\hline
$id$ & $1$ & $3$ & $4$\\
\hline
$predecessor$ & $\diamondsuit$ & $1$ & $3$ \\
\hline
$successor$ & $3$ & $4$ & $\diamondsuit$ \\
\hline
\end{tabular}
$\xrightarrow{N_{2} \textsc{Join}}$
\begin{tabular}{|c|c|c|c|c|}
\hline
\multicolumn{5}{|c|}{\boldmath{$S_1$}} \\
\hline
$id$ & $1$ & \boldmath{$2$} & $3$ & $4$\\
\hline
$predecessor$ & $\diamondsuit$ & \boldmath{$undef$} & $1$ & $3$ \\
\hline
$successor$ & $3$ & \boldmath{$3$} &  $4$ & $\diamondsuit$ \\
\hline
\end{tabular}
\vspace{1mm}
$\xrightarrow{N_{3} \textsc{FairLeave}}$
\begin{tabular}{|c|c|c|c|}
\hline
\multicolumn{4}{|c|}{\boldmath{$S_2$}} \\
\hline
$id$ & $1$ & $2$ & $4$\\
\hline
$predecessor$ & $\diamondsuit$ & $undef$ & $1$ \\
\hline
$successor$ & $4$ & $3$ & $\diamondsuit$ \\
\hline
\end{tabular}
\end{center}

A similar example can be given for \textsc{UnfairLeave}. \hfill $\square$

\end{example}

In the sequel, we show that a stable pair of nodes in a Chord
network, which executes a regular run, eventually becomes stable
after adding/removing of a node between them (the theorems
\ref{succend}-\ref{chord_cycle}). Corollary \ref{cor4} formulates
the corresponding statement for a stable network. Finally, we
prove that the proposed key-handling correctly distributes keys
and answers queries (Theorem \ref{golden_rule} and Corollary
\ref{corol1}).

The first theorem expresses
% the correctness of the
that the rule \textsc{FindSuccessor} will terminate in a finite
number of steps.
%, that the result will always be a valid node and
% that result will be unique).
It corresponds to Theorem IV.2
from\cite{Chord,Chord-TR,Chord-IEEE}.

%The first three theorems express the correctness of the
%\textsc{FindSuccessor} rule (that will terminates in a finite
%number of steps, that the result will always be a valid node and
%that result will be unique).
%
%These theorems \ref{succend} -- \ref{fsthv2} correspond
%to Theorem IV.2 and IV.4 from \cite{Chord,Chord-TR,Chord-IEEE}.

\begin{theorem}\label{succend}
Let $n \in Chord$ be the node which fires the rule
\textsc{FindSuccessor} for $h \in \{0, 1, \ldots , N-1\}$. Let
$m'$ be the minimal element of $Chord$ such that $h \leqslant m'$.
If the pair $\langle n, m' \rangle$ is stable in that state, the
node $n$ will get the result after a finite number of moves.

\end{theorem}

Theorems \ref{stabthj} -- \ref{chord_cycle} guarantee that the successor and predecessor pointers for each node will be eventually
up to date after a node joins, or unfair leaves the network. In the corresponding proofs we will use some finite initial sequences of
runs. Due to the fact that the \textsc{Stabilize} and \textsc{UpdatePredecessor} are applied periodically by all nodes in a network, we will
mention only those applications which change the values of the functions $predecessor$ and $successor$.

Note that, in each proof we will consider some fixed linearization
of moves, but according to Corollary \ref{cor1}, all
linearizations of the corresponding regular run will result in the
same final state.

Theorem \ref{concjoin} corresponds to Theorem IV.3 from \cite{Chord,Chord-TR,Chord-IEEE}.

\begin{theorem}\label{stabthj}
Let a peer join a Chord network, between two nodes which constitute a stable pair. Then, there is a number $k>0$ of steps, such that if no other join rule
happens in the meantime, the \textsc{Stabilize} rule will bring the starting pair to be stable after $k$ steps.

\end{theorem}

\begin{theorem}[Concurrent joins]\label{concjoin}
Let a Chord network contain a stable pair. If a sequence of \textsc{Join} rules is executed between the nodes which form this stable pair, interleaved with \textsc{Stabilize}, \textsc{UpdatePredecessor}
and \textsc{Update\_fingers}, then there is a number $k > 0$ of steps, such that after the last \textsc{Join} rule, the starting pair of nodes will be
stable after $k$ steps.

\end{theorem}

\begin{theorem}\label{stabthl}
Let a Chord network contain a stable pair and let a node between them leave the network.
% in an unfair way, and break the ring of the successors pointers.
Then, there is a number $k \geqslant 0$ of steps, such that if no
\textsc{Join} rule happens at the considered part of the network
in the meantime,
% $Stabilize$ and $Update\_predecessor$ will rebuild
% the ring of successor pointers will be rebuild after $k$ steps.
the pair will be brought into a stable state after $k$ steps.
\end{theorem}

\begin{theorem}\label{stab_mix_join_leave}
Let a Chord network contain a stable pair. Let a node which is
between those nodes leave the network following by several nodes
which want to join between them.
% in an unfair way, and break the ring of the successors pointers.
Then, there is a number $k \geqslant 0$ of steps, such that
% the ring of successor pointers will be rebuild after $k$ steps.
the considered pair will be brought into a stable state after $k$
steps.
\end{theorem}

Note that the restriction from the formulation of Theorem \ref{stab_mix_join_leave}, that no other leave-rules are allowed after the
first one, is not essential. According to the definition of regular runs, leave-rules can be executed only between nodes which constitute a stable pair, and we
can consider an execution of a sequence of join rules interleaved with leave-rules, and obtain the same result. The above statement
will hold for each subsequence which starts with a leave rule followed by several join rules. Thus, we have the following:

%%%%%%%

\begin{corollary} \label{cor3}
Let a Chord network contain a stable pair. Let a node, which is in between those nodes, leave the network.
% in an unfair way, and break the ring of the successors pointers.
Then, there is a number $k \geqslant 0$, such that
% such that the ring of successor pointers will be rebuild after $k$ steps.
the considered pair of nodes will become stable after $k$
moves.
\end{corollary}

%%%%%%

Theorem \ref{chord_cycle} incorporates all previous ideas, and is the main statement concerning correctness of maintaining topological
structure of Chord networks.

\begin{theorem}\label{chord_cycle}
Let a finite initial segment of a run produce the state $S$ of a
Chord network. Then, for every pair of nodes $n, n' \in Chord$,
there is a number $k \geqslant 0$, such that
% the ring of successor pointers will be rebuild after k steps.
$\langle n, n' \rangle$ will become stable after $k$ moves.
\end{theorem}

Since a network is stable in a state if all pairs of nodes from
the network are stable in that state, we have:

\begin{corollary} \label{cor4}
Let a finite initial segment of a run produce the state $S$ of a
Chord network. Then, there is a number $k \geqslant 0$, such that
% the ring of successor pointers will be rebuild after k steps.
the network will become stable after $k$ moves.
\end{corollary}

Finally, the next two statements say that our formalization
consistently manipulates distributed keys. Theorem
\ref{golden_rule} states that $(key, value)$ pairs are properly
distributed over the network. Informally, it follows from the
facts that for every $n \in Chord$, $hash(key) \leqslant n$ for
the keys for which $n$ is responsible for, and that all rules that
manipulate $(key, value)$ pairs invoke \textsc{FindSuccessor}
rule.

\begin{theorem}[Golden rule]\label{golden_rule}
$$
\forall((key, value) \in Keys \times Values, n \in Chord) ((key, value) \in keyvalue(n)
$$
$$
\Rightarrow member\_of(hash(key), predecessor(n),n).
$$
\end{theorem}

Corollary \ref{corol1} follows from the definition of
\textsc{Get}, and the theorems \ref{succend} and
\ref{golden_rule}:

\begin{corollary} \label{corol1}
If \textsc{Get} returns $undef$ for some $key \! \in \! Keys$, then there is no $value \! \in \! Values$ such that $(key, value)$ pair is
stored in the Chord network.
\end{corollary}
Namely, according to Theorem \ref{golden_rule}, all $(key, value)$
pairs are stored properly, and from Theorem \ref{succend}
\textsc{Get} considers only the $(key, value)$ pairs stored in the node $N$ which
satisfy condition $member\_of($ $hash(key), predecessor(id(N)),id(N))$.

\section{Conclusion}

\label{conclusion}

In this paper we have presented an ASM-based formalization of the Chord protocol. We have proved that the proposed formalization is
correct with respect to the regular runs. Up to our knowledge, it is the first comprehensive formal analysis of Chord presented in the
literature which concerns both maintenance of the ring topology and data distribution. We have also indicated that if we consider all
possible runs, incorrect behavior of Chord protocol could appear.

Possible direction for further work is to apply similar technique to describe other DHT protocols. For example, an interesting
candidate for examination in the ASM-framework could be Synapse, a protocol for information retrieval over the inter-connection of
heterogeneous overlay networks defined in \cite{LTVBCM10}, and applied in \cite{ohrid}.

Another challenge could be verification of the given description in one of the formal proof assistants (e.g., Coq, Isabelle/HOL). It
might also produce a certified program implementation from the proof of correctness of our ASM-based specification.

\bibliographystyle{fundam}

\end{document}